\documentstyle[sprocl]{article}

\input{psfig.sty}

\def\be{\begin{equation}}
\def\ee{\end{equation}}
\def\bea{\begin{eqnarray}}
\def\eea{\end{eqnarray}}
\begin{document}

\title{The Sphaleron Rate:  Where We Stand}

\author{Guy D. Moore}

\address{Dept.~of Physics, McGill University, 3600 University St. \\
        Montreal, Qc H3A 2T8 Canada } 


\maketitle\abstracts{I review what we know about the ``sphaleron rate'',
which is the efficiency of baryon number violation at high temperatures
$T \sim 100 \: {\rm GeV}$ in the Standard Model.  The leading behavior
at weak coupling in the symmetric phase is known accurately;
$\Gamma = (10.7 \pm .7) (g^2 T^2 / m_D^2) \log(m_D/g^2T)\alpha_w^5 T^4$.
At realistic values of the coupling our accuracy is worse.  We also now
have the tools to determine the rate nonperturbatively in the broken
electroweak phase; the sphaleron rate there is slower than
perturbative estimates.}

\section{Introduction}

The Universe is filled with matter, and virtually no antimatter.  This
``unusual'' situation can only be explained without appealing to the
initial conditions of the Universe if, at some early epoch in the
history of the universe, baryon number was not a conserved quantity.

As a matter of fact, baryon number is {\em not} a conserved quantity in
the standard model \cite{tHooft}.  
Furthermore, while its violation under ordinary conditions is pitifully
inefficient, that ceases to be true at very high temperatures, of order
the weak scale, where electroweak symmetry is restored.  These facts are
the backdrop for the subject of electroweak baryogenesis, which attempts
to use them to explain why the baryon number density of the current
universe is what it is.

In this talk I will not discuss baryogenesis.  Rather my emphasis is on
strengthening its foundations by investigating more accurately exactly
how efficiently baryon number is violated under hot conditions.  Besides
the obvious application to the study of baryogenesis, this is also a
useful thing to do because it forces us to develop tools for dealing
with the infrared physics of hot Yang-Mills theory.  Also it is the only
part of a baryogenesis calculation which is generic to all extensions of
the standard model, because it only depends on the gauge group, and only
very weakly on the Higgs sector.  In fact, for much of the talk I will
neglect the Higgs fields altogether and just study Yang-Mills theory.

\subsection{What we want to know}

Before going any further I should fix notation and explain what I will
measure.  The anomaly equation relates baryon number to the Chern-Simons
number of the SU(2) weak fields, through
\begin{equation}
\frac{1}{3} N_{\rm B}(t) = N_{\rm CS} \equiv \frac{1}{8 \pi^2} \int^t dt' 
	\int d^3 x E_i^a B_i^a(x,t') \, ,
\label{define_NCS}
\end{equation}
where $E$ and $B$ are the SU(2) electric and magnetic fields, and I
normalize so the gauge field $A$ has units of inverse length and $g^2$
appears in the denominator in the Lagrangian.  The constant of
integration from the indefinite time integral is fixed by requiring
$N_{\rm CS}$ to be an integer for a vacuum configuration.
$N_{\rm CS}$ is called the Chern-Simons number.
Because magnetic fields are always transverse (Gauss' Law for
magnetism), the evolution of Chern-Simons number depends on the physics
of the transverse sector.

Having defined Chern-Simons number I can define its diffusion constant,
\begin{equation}
\label{define_Gamma}
\Gamma \equiv \lim_{V \rightarrow \infty} \lim_{t \rightarrow \infty}
	\frac{ \langle (N_{\rm CS}(t) - N_{\rm CS}(0))^2 \rangle}{Vt} \,
	,
\end{equation}
where the angular brackets $\langle \rangle$ mean an average is taken
over the thermal density matrix.  $\Gamma$
is often referred to as the ``sphaleron rate.''  The reason we
care about it is that there is a fluctuation dissipation relation
between it and the relaxation rate for a chemical 
potential for baryon number.  I will not discuss this in detail, see
instead \cite{KhlebShap,Mottola,RubShap}.
I also comment that for $N_{\rm CS}$ to diffuse requires nonperturbative
physics, and nonperturbative physics is only unsuppressed, at high
temperatures and weak coupling, on length scales $\geq (1 / g^2 T) \;$
\cite{ArnoldMcLerran}.

It is interesting to know $\Gamma$ in two regimes.  The first is the
electroweak symmetric phase.  Almost all baryogenesis mechanisms will
give a final baryon number directly proportional to its value here.  For
this reason we would like to know it with some accuracy here, which
makes the calculation tricky.  The other
regime where we want to know $\Gamma$ is in the broken electroweak
phase, {\em immediately} after the electroweak phase transition, which
is to say, right after the baryons were allegedly produced.  In this
case what we want to know is, are the baryons safe, or will they
subsequently be destroyed?  Here the strength of the phase 
transition is important; the real question is, how strong must the phase
transition be to prevent the baryons from getting destroyed?  To answer
this question with descent resolution we only need to know $\Gamma$ to
within $\pm 1$ in the exponent; however $\Gamma$ is exponentially small
and perturbation theory is not yet reliable, which will make this
calculation tricky as well.

\subsection{Approximations I will need}

Determining $\Gamma$ requires determining unequal Minkowski time
correlators at finite temperature in a quantum field theory.
Furthermore, if the answer is to be nontrivial the field theory must be
showing nonperturbative physics.  No one knows how to do this directly.
Therefore I will obviously need to make some approximations.

What saves us is that the SU(2) sector is weakly coupled.  This allows
two key approximations which make the problem tractable.  First, the
infrared behavior of the theory is classical up to parametrically
suppressed corrections.  Second, the ultraviolet behavior of the theory
is perturbative.  Here, infrared means $k \ll \pi T$, while ultraviolet
means $k \gg \alpha_w T$.  When the coupling really is weak, there is an
overlap between these two regimes, and every degree of freedom can be
treated with one approximation or the other.  Then, one can integrate out
those degrees of freedom which are perturbative, and treat the
remaining, classical theory nonperturbatively on the lattice.

In what follows I will first discuss what we learn by treating
perturbatively and integrating out everything we can.  Then I will step
back and only integrate out the highest $k$ modes analytically, leaving a
larger and more inclusive theory for numerical work.  Finally I discuss
what we can do in the broken phase.  The complete details for these
three approaches can be found in \cite{Bodek_paper}, \cite{particles},
and \cite{broken_nonpert} respectively.

\section{Leading log}

Dietrich B\"{o}deker has shown that it is possible to integrate out all
degrees of freedom with momentum scale $k \geq g^2 T \log(1/g)$, and
that doing so produces an effective theory for the remaining $k \sim g^2
T$ degrees of freedom which is classical Yang-Mills theory under
Langevin dynamics \cite{Dietrich}.  The physical origin of this
effective theory has been
discussed by Arnold's group \cite{ASY2}.  Integrating out the
modes with $k \geq gT$ gives the well known hard thermal loop effective
theory \cite{Pisarski}.  The behavior of the infrared modes in this
theory is overdamped \cite{ASY}, 
which just follows from Lenz's law and the fact
that the plasma is highly conducting.
The conductivity is $k$ dependent on scales shorter than some mean
collision length, which in an abelian theory is the large angle
scattering length.  However in a nonabelian theory a particle's charge
is changed by scattering.  The mean length for a particle to travel
before its charge is randomized is
\begin{equation}
l_{\rm scatt}^{-1} = \frac{g^2 T}{2 \pi} \left[ \log \frac{m_D}{g^2 T}
	+ O(1) \right] \, ,
\end{equation}
so on scales longer than this the strength of damping is $k$
independent.  Therefore, in the approximation that the scale $1/g^2 T$
is well separated from the scale $1 / (g^2 T \log(1/g))$, the infrared
fields obey Langevin dynamics on long time scales.

Langevin dynamics have two nice features.  First, the Langevin dynamics
of 3-D Yang-Mills theory are free of UV problems, and a zero lattice
spacing limit exists.  Second, Langevin dynamics are very easy to
put on the lattice.  Therefore the emphasis should be on controlling
systematics, such as
\begin{enumerate}
\item the thermodynamic match between lattice and continuum,
\item the match between lattice and continuum Langevin time scales,
\item topological definition of $N_{\rm CS}$, and
\item the large volume and long time limits.
\end{enumerate}

All of these systematics can be controlled.  The first is discussed in
\cite{Oapaper}, the second in \cite{Bodek_paper}, and the third in
\cite{broken_nonpert}.  A volume $8 / g^2 T$ on a side is large enough
to achieve the large volume limit, so I use a volume $16/g^2 T$ on a
side for overkill.  The failure to achieve the large time limit is
reflected in the statistical error bars.

The results, which show
beautiful lattice spacing independence, are presented in Table
\ref{bodek_table}, which gives the coefficient $\kappa'$ for the
equation 
\begin{equation}
\Gamma = \kappa' \left( \log \frac{m_D}{g^2 T} + O(1) \right) \left( 
	\frac{g^2 T^2}{m_D^2} \right) \alpha_w^5 T^4 \, .
\label{leadinglog}
\end{equation}
These results settle the question, ``What is the sphaleron rate in the
symmetric phase, in the extreme weak coupling limit?''

\begin{table}[t]
{\centerline{\mbox{
\begin{tabular}{|c|c|c|c|} \hline
lattice spacing $a$ & Volume & Langevin time &
        $\kappa' \pm$ statistical error \\ \hline
$2/3g^2T$  &  $(  8 / g^2 T)^3$ & $290000a^2$ & $10.44 \pm 0.23$ \\ \hline
$2/3g^2T$  &  $( 16 / g^2 T)^3$ & $49500 a^2$ & $10.30 \pm 0.21$ \\ \hline
$2/5g^2T$  &  $( 16 / g^2 T)^3$ & $21000 a^2$ & $10.70 \pm 0.67$ \\ \hline
$2/7g^2T$  &  $( 16 / g^2 T)^3$ & $42000 a^2$ & $10.25 \pm 0.79$ \\ \hline
\end{tabular}}}}
\caption{\label{bodek_table}
        Results for $\kappa'$ at three lattice spacings and two lattice
volumes.  The results show excellent spacing and volume independence.}
\end{table}

\section{Beyond the leading log}

In the last section I determined the coefficient of the leading log
behavior with very good precision.  The problem is the $(+ O(1))$
appearing in Eq. (\ref{leadinglog}).  How well does the expansion in
$\log (1/g)$ converge?

The answer is, probably very poorly.  To see this, compare the free path
for color randomization, $l_{\rm max} = 2 \pi/g^2 T \log(1/g)$, 
to the size of a
box for which the large volume limit has already been reached, $8 / g^2
T$.  There is not a large separation between these scales.  In fact, it
is not clear whether $l_{\rm max}$ is smaller than the
scale characterizing nonperturbative physics, which must after all be
well shorter than the dimension of a box which shows large volume
behavior.  The problem is that B\"{o}deker's approach
requires integrating out modes for which the perturbative treatment may
not be very reliable.  To test this, and to try to determine the
sphaleron rate beyond leading log, we need to integrate out less, and
make the numerical model include the $gT$ as well as $g^2 T$ scales.

An effective action
for the theory with the $k \sim T$ modes integrated out is known
\cite{HTL_action}, and goes by the name of the hard thermal loop (HTL)
effective action.  It is nonlocal, which is not surprising, since its
construction involves integrating out propagating degrees of freedom
in a Minkowski theory.  Unfortunately nonlocality is very problematic
for a numerical implementation.

\begin{figure}[t]
\centerline{\psfig{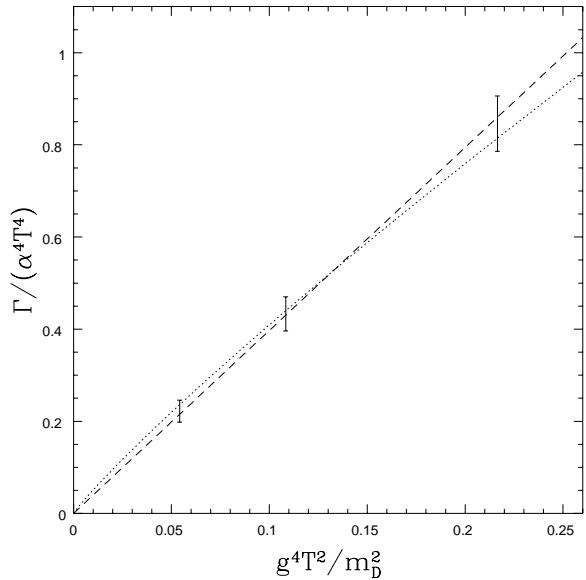}}
\caption{\label{fig_particles} Sphaleron rate plotted against inverse
HTL strength, for the lattice theory plus ``particles'' used to generate
the hard thermal loops.  The dashed fit is a straight line, while the
dotted fit incorporates the known logarithmic dependence on $m_D$ found
in the previous section.}
\end{figure}

A solution to this problem was proposed
a few years ago by Hu and M\"{u}ller \cite{HuMuller}.  The idea is that,
rather than add the HTL action itself, one adds a set of local degrees
of freedom which, if integrated out, would also yield the HTL effective
action.  Since the HTL action represents the propagation of a set of
high momentum, charged particles, what we add are a bunch of high
momentum, charged classical particles.  For the idea to work it is
necessary to add particles in such a way that the numerical model
retains an exact gauge invariance, and possesses a conserved energy and
phase space measure so that thermodynamic averages are well defined.  We
present an implementation which satisfies these conditions in
\cite{particles}, and refer the reader there for the (quite complicated)
details. 

With a model which reproduces the HTL resummed IR physics in hand, it is
possible to test Arnold, Son, and Yaffe's claim that the IR dynamics are
overdamped, directly.  The results are shown in Figure
\ref{fig_particles}, which shows that $\Gamma$ does vanish linearly as
$m_D^2$ is increased.  The data are not good enough to show
unambiguously whether or not B\"{o}deker's log is present.  Fitting the
data assuming it is, the $(+O(1))$ in Eq. (\ref{leadinglog}) 
turns out to be about
3.6, which indicates that the expansion in $\log(m_D/g^2 T)$ is not a
very good one.  For the physical value of $m_D^2 = (11/6) g^2 T^2$ and
$g^2 \sim 0.4$, the sphaleron rate is $(20 - 25) \alpha^5 T^4$, with
systematics dominated errors of the order of $30\%$.  The main remaining
problems to be addressed involve lattice spacing effects and the
interactions of the added ``particle'' degrees of freedom with the most
UV lattice modes.

\section{The broken electroweak phase}

In the previous two sections the approach has been to find a numerical
system which has the same physics as thermal Yang-Mills theory, and then
to evolve it and measure directly the correlator,
Eq. (\ref{define_Gamma}), which tells how efficiently baryon number is
violated.  However, this approach fails completely in the broken
electroweak phase, because the rate of topological transitions is 
so small that no reasonable amount of numerical evolution would see any
transitions at all.  Another alternative, perturbation theory, is not
very reliable close to the electroweak phase transition.  We know for
instance that the one loop and two loop effective potentials give quite
different answers for the strength of the transition, and no one knows
how to compute the sphaleron rate beyond the one loop level.  Some other
technique, nonperturbative but not strictly real time, is needed.

\begin{figure}[t]
\centerline{\psfig{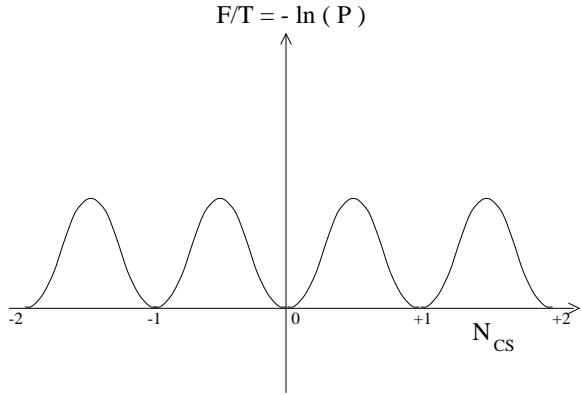}}
\caption{\label{cartoon}  ``Cartoon'' of the free energy dependence on
$N_{\rm CS}$.}
\end{figure}

The reason for the suppression of the sphaleron rate in the broken
phase is shown, in cartoon form, in Figure \ref{cartoon}.  There is a
free energy barrier between minima, meaning that almost none of the
weight of the thermal ensemble lies in states intermediate between
vacua.  The figure also suggests how I will determine the sphaleron
rate in the broken phase.  I should define $N_{\rm CS}$ (or some
appropriate observable) on the lattice, and measure how the free energy
depends on it.  This is not enough to give the real time rate, but with
some more work one can turn the height of the barrier into the real time
rate.  

\subsection{Defining $N_{\rm CS}$}

I begin by defining Chern-Simons number on the lattice.  The obvious
approach is to use the same definition as in the continuum,
Eq. (\ref{define_NCS}).  Note that the integral over ``time'' in that
equation could really be an integral along any path through the space of
configurations, not just one generated by Hamilton's equations.  In
particular one can fix the constant of integration by having the path
begin or end at a vacuum configuration.  

There is a problem on the lattice, which is that no lattice
implementation of $E_i^a B_i^a$ is exactly a total derivative.
Therefore $N_{\rm CS}$ defined through Eq. (\ref{define_NCS}) and
implemented on the lattice would depend on the path chosen.  The
resolution is to choose a unique and particularly sensible path,
the gradient flow (cooling) path, that is, the path through
configuration space along which the energy falls most rapidly.  In
continuum notation the path (parameterized by a cooling time $\tau$) is
given by
\begin{equation}
\frac{dA(x,\tau)}{d\tau} = - \frac{\partial H}{\partial A(x)} \, ,
\end{equation}
where $H$ is the Hamiltonian and all indices have been suppressed.
Besides being unique, this path also has the benefit that it moves
quickly towards configurations where the gauge fields are smooth.  This
minimizes the impact of lattice artifacts, which were for instance
responsible for $E_i^a B_i^a$ not being a total derivative.  Further,
the path automatically goes to a vacuum configuration.

This definition of $N_{\rm CS}$ has two problems, both easily resolved.
First, $N_{\rm CS}$ is UV poorly behaved.  For instance, its mean
squared value diverges as $V/a$, with $V$ the physical volume and $a$
the lattice spacing (or other regulator).  This is resolved by using not
$N_{\rm CS}$ but the Chern-Simons number of a configuration after an
initial length $\tau_0$ of gradient flow.  Our measurable is then
dependent on an unphysical parameter $\tau_0$, but it is UV finite, and
physical measurables such as $\Gamma$ will be $\tau_0$ independent in
the end.

The second problem is that performing gradient flow down to the vacuum
is intensely numerically expensive; yet we will need to do so thousands
of times to determine the free energy distribution.  This problem is
solved by blocking.  Gradient flow destroys information, and in
particular it destroys almost all the UV information; so nothing is lost
by blocking after some modest amount of gradient flow.  The numerical
savings are immense, and (if we use an $O(a^2)$ improved lattice
Hamiltonian and implementation of $E_i^a B_i^a$) almost no accuracy is
lost.  

\begin{figure}[t]
\centerline{\psfig{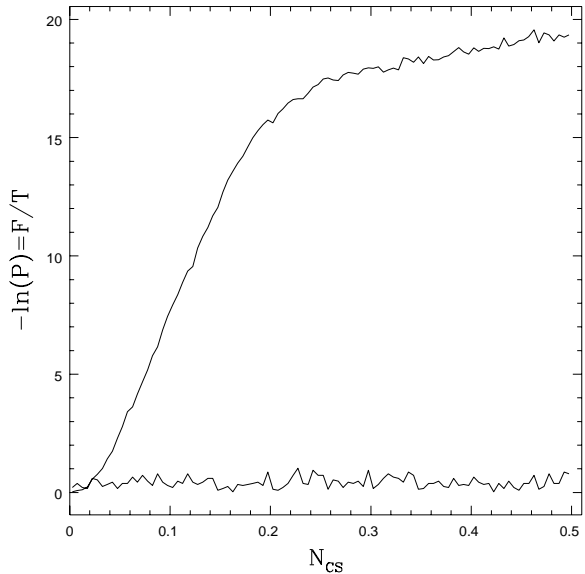}}
\caption{\label{Free_energy} Free energy as a function of $N_{\rm CS}$,
at the critical temperature in a $(16/g^2 T)^3$ cubic box and using
$\tau_0 = 3.6/(g^2T)^2$, when the ratio of the Higgs self-coupling to
the gauge coupling was $\lambda/g^2 = 0.039$.  The upper curve is the
broken phase value and the lower curve is the symmetric phase value.}
\end{figure}

Using this definition of $N_{\rm CS}$, it is possible to determine the
free energy (probability distribution) as a function of $N_{\rm CS}$ by
standard multicanonical Monte-Carlo techniques.
A sample result is shown in Figure \ref{Free_energy}.

\subsection{Turning probabilities into rates}

One cannot read off the sphaleron rate from Figure \ref{Free_energy}
alone; in fact the height of the barrier in the figure depends on an
arbitrary parameter $\tau_0$.  To get $\Gamma$ from the figure we need
to know
\begin{equation}
\langle \dot{N} \rangle \equiv \left\langle \left| 
	\frac{dN_{\rm CS}}{dt} \right|_{N_{\rm CS} = 0.5} 
	\right\rangle \, ,
\label{def_Ndot}
\end{equation}
the mean rate at which $N_{\rm CS}$ is changing during a crossing of the
barrier.  Multiplying the probability density at the top of the barrier
by $\langle \dot{N} \rangle$ turns the probability density into a
probability flux per unit time.  

It is straightforward to measure $\langle \dot{N} \rangle$ numerically.
First, we use multicanonical means to get a sample of configurations
with $N_{\rm CS} \simeq 0.5$.  Then, for each we draw momenta randomly
from the thermal ensemble and perform a very short period of Hamiltonian
evolution, measuring $N_{\rm CS}$ before and after.  Then $|dN_{\rm
CS}/dt|$ is approximated by $|N_{\rm CS}(0) - N_{\rm CS}(\delta
t)|/(\delta t)$, and $\langle \dot{N} \rangle$ is the average of this
over the sample.  Also, one must divide by the volume used
in the lattice simulation, to convert the rate of topological
transitions to the rate per unit volume.

However, $\Gamma$ does not equal the probability flux per unit time over
the barrier; it is how often one goes from being in one topological
vacuum to being in another.  It is possible to cross the barrier several
times on the way from one minimum to its neighbor, or to cross an even
number of times and return to the starting vacuum.  This leads to a
correction called the ``dynamical prefactor,'' which is the ratio of
true topological vacuum changes to crossings of the top of the barrier.
To compute it, we use multicanonical means to get a sample of $N_{\rm
CS} = 0.5$ configurations.  Then each is evolved under Hamiltonian
dynamics, both forward and backwards in time, until it settles in a
topological vacuum.  The dynamical prefactor is 
\begin{equation}
{\rm Prefactor} = \sum_{\rm sample} \frac{1}
	{\# {\rm \; crossings}} (\Delta N_{\rm CS})^2 \, ,
\end{equation}
where $\Delta N_{\rm CS}$ is the difference in $N_{\rm CS}$ between the
starting and ending vacua.  It is $\pm 1$ if there were and odd number of
$N_{\rm CS}=0.5$ crossings and $0$ if there were an even number; we
never observe prompt crossings from one topological minimum to another
which is not its immediate neighbor.

The hard thermal loops appear in the dynamical prefactor, which is
parametrically of order $(g^4 T^2/m_D^2) \log(m_D/g^2 T)$.  However,
using the techniques of the last section to include the HTL effects in
the calculation of the prefactor shows that, for realistic values of
$m_D^2$, the importance of HTL's is weak.  This is expected, or at least
we should expect that the dependence is weaker than in the symmetric
phase, because broken phase baryon number violation should be mediated
by a spatially smaller configuration, involving higher frequency modes
which are less overdamped.

\begin{figure}
\centerline{\psfig{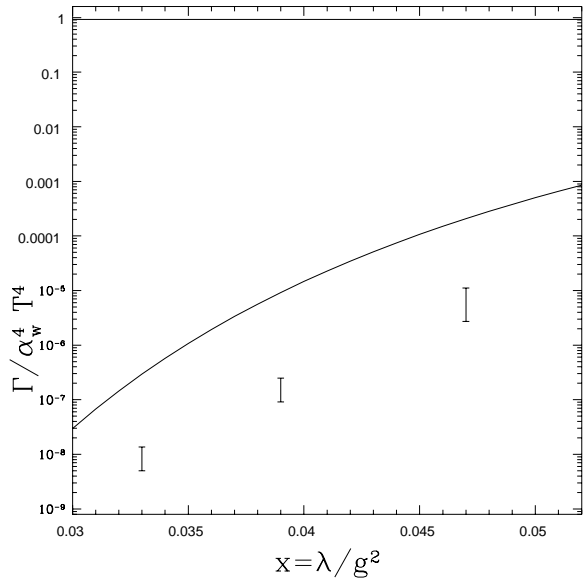}}
\caption{\label{final_result} Chern-Simons number diffusion constant
$\Gamma$ immediately after the electroweak phase transition as a
function of $\lambda / g^2$, which at tree level equals $m_H^2 / 8
m_W^2$.  The solid line is a perturbative estimate, the line at the top
of the figure is the symmetric phase rate.}
\end{figure}

The final result for $\Gamma$ in the broken electroweak phase, at the
electroweak phase transition temperature and for a range of scalar
self-couplings, is plotted in Figure \ref{final_result}, which also
compares it to a perturbative result based on a two loop potential and
the zero mode calculations from Carson and McLerran \cite{Carson}.  The
actual rate is substantially but not drastically slower than the
perturbative estimate.  For comparison, the value needed to avoid baryon
number washout after the transition, in the standard cosmology, is
$\Gamma \sim 10^{-7} \alpha_w^4 T^4$.

\section{Conclusion}

Tools now exist to calculate the baryon number violation rate in
both the symmetric and broken electroweak phases.  In the symmetric
phase the rate behaves parametrically as $\alpha_w^5$, with a 
logarithmic correction found by B\"{o}deker which is numerically
small.  The rate at a realistic $m_D^2$ is $20-25 
\alpha_w^5 T^4$, with systematic errors, estimated to be of order
$30\%$, dominating statistical errors.  In the broken phase the 
rate is smaller than a perturbative estimate, but still too large to
save baryogenesis in the minimal standard model.

\section*{Acknowledgments}
I would like to thank Peter Arnold, Dietrich B\"{o}deker, Dam Son, and
Larry Yaffe, who have consistently shared results before publication and
with whom I have had many useful discussions.


\section*{References}
\begin{thebibliography}{99}
\bibitem{tHooft} G. t'Hooft, Phys. Rev. Lett. {\bf 37},8 (1976).
\bibitem{KhlebShap} S. Khlebnikov and M. Shaposhnikov, Nucl. Phys.
        {\bf B308}, 885 (1988).
\bibitem{Mottola} E. Mottola and S. Raby, Phys. Rev. {\bf D 42}, 
	4202 (1990).
\bibitem{RubShap}  V. Rubakov and M. Shaposhnikov,
	Phys. Usp. {\bf 39}, 461 (1996)
	[Usp. Fiz. Nauk {\bf 166}, 493 (1996)].
\bibitem{ArnoldMcLerran} P. Arnold and L. McLerran, Phys. Rev. {\bf D 36},
        581 (1987).
\bibitem{Bodek_paper} G. D. Moore, hep-ph/9810313.
\bibitem{particles} G. D. Moore, C. Hu, and B. M\"{u}ller, Phys. Rev. 
	{\bf D 58}, 045001 (1998).
\bibitem{broken_nonpert} G. D. Moore, Phys. Rev. {\bf D 59}, 
	014503 (1999).
\bibitem{Dietrich} D. B\"{o}deker, Phys. Lett. {\bf B 426}, 351 (1998).
\bibitem{ASY2} P. Arnold, D. Son, and L. Yaffe, 
	UW/PT 98-10, hep-ph/9810216.
\bibitem{Pisarski} E. Braaten and R. Pisarski, Nucl. Phys. {\bf B 337}, 
	569 (1990).
\bibitem{ASY} P. Arnold, D. Son, and L. Yaffe, Phys. Rev.
	{\bf D 55}, 6264 (1997).
\bibitem{Oapaper} G. D. Moore, Nucl. Phys. {\bf B 493}, 439 (1997);
	Nucl. Phys. {\bf B 523}, 569 (1998).
\bibitem{HTL_action} J.C. Taylor and S.M.H. Wong, 
        Nucl. Phys. {\bf B 346}, 115 (1990).
\bibitem{HuMuller} C. Hu and B. M{\"u}ller,
 	Phys. Lett. {\bf B409}, 377 (1997).
\bibitem{Carson} L. Carson and L. McLerran, Phys. Rev. {\bf D 41}
	(1990) 647.
\end{thebibliography}
\end{document}